\documentclass[%
 reprint,
 amsmath,amssymb,
 aps,
]{revtex4-2}

\usepackage[utf8]{inputenc}
\usepackage[english]{babel}

\usepackage{amsthm,amsmath,amsfonts,amssymb}
\usepackage{braket,nicefrac}
\usepackage{arydshln} 

\PassOptionsToPackage{breaklinks}{hyperref}
\usepackage[colorlinks=true,linkcolor=blue,urlcolor=blue,citecolor=blue,anchorcolor=green,pdfusetitle]{hyperref}
\usepackage{bbm,dsfont,xspace,xparse}
\usepackage{lmodern,newtxtt,microtype}
\usepackage{tabularx,xifthen,afterpage,booktabs}
\usepackage{qcircuit,algpseudocode,algorithm}
\usepackage{xurl}

\usepackage{microtype} 
\usepackage[dvipsnames]{xcolor}
\usepackage{upgreek}

\usepackage{cleveref}
\crefname{lemma}{lemma}{lemmas}
\crefname{proposition}{proposition}{propositions}
\crefname{definition}{definition}{definitions}
\crefname{theorem}{theorem}{theorems}
\crefname{corollary}{corollary}{corollaries}
\crefname{section}{section}{sections}
\crefname{appendix}{appendix}{appendices}
\crefname{figure}{fig.}{figs.}
\crefname{equation}{eq.}{eqs.}
\crefname{table}{table}{tables}
\newtheorem{theorem}{Theorem}
\newtheorem{definition}[theorem]{Definition}

\newtheorem{lemma}[theorem]{Lemma}

\usepackage{listings}
\newcommand{\C}{{\mathbb{C}}} 

\renewcommand{\O}{\mathcal{O}}
\newcommand{\id}{{\mathbbm{1}}} 
\newcommand{\Exp}[2][]{\mathbb{E}{%
	\ifthenelse{\isempty{#1}}{}{_{#1}}}%
	\left[{#2}\right]} 
\newcommand{\dx}[1][]{\mathrm{d}{\ifthenelse{\isempty{#1}}{}{#1}}} 

\newcommand{\norm}[1]{\|{#1}\|}

\newcommand{\Abs}[1]{\left|{#1}\right|}
\newcommand{\Tr}[2][]{\mathop{}\text{Tr}{%
	\ifthenelse{\isempty{#1}}{}{_{#1}}}%
	\left({#2}\right)}

\newcommand{\ketbra}[2]{\ket{#1}\!\!\bra{#2}}
\newcommand{\proj}[1]{\ketbra{{#1}}{{#1}}}

\newcommand{\cc}[1]{\mathsf{{#1}}}	

\definecolor{redrev}{rgb}{0.77, 0.01, 0.2}

\newcommand{\parahead}[1]{\noindent\textbf{{#1}}} 

{%
}%

\begin{document}

\title{Quantum algorithm for estimating $\alpha$-Renyi entropies of quantum states}

\author{Sathyawageeswar Subramanian}
\email{ss2310@cam.ac.uk}
\affiliation{CQIF, DAMTP
  University of Cambridge
  Cambridge CB3 0WA}
\author{Min-Hsiu Hsieh}%
 \email{min-hsiu.hsieh@uts.edu.au}
\affiliation{Centre for Quantum Software and Information
  Faculty of Engineering and Information Technology
  University of Technology Sydney, Australia
  }%

\begin{abstract}
We describe a quantum algorithm to estimate the $\alpha$-Renyi entropy of an unknown density matrix $\rho\in\C^{d\times d}$ for $\alpha\neq 1$ by combining the recent technique of quantum singular value transformations with the method of estimating normalised traces in the one clean qubit model. We consider an oracular input model where the input state is prepared via a quantum oracle that outputs a purified version of the state, assumed to be non-singular. Our method outputs an estimate of the $\alpha$-Renyi entropy to additive precision $\epsilon$, using an expected total number $\O(\nicefrac{1}{(x\epsilon)^2})$ of independent applications of a quantum circuit which coherently queries the input unitary $\O(\nicefrac{1}{\delta}\log\nicefrac{d}{\epsilon})$ times, in each case measuring a single output qubit. Here $\delta$ is a lower cutoff on the smallest eigenvalue of $\rho$ and $x=\frac{1}{d}\!\Tr{\rho^\alpha}$. The expected number of measurements made in this method can be compared to results in the sample complexity model that generally require $\Theta(d^2/\epsilon^2)$ samples. Furthermore, we also show that  multiplicative approximations can be obtained by iteratively using additive approximations, with an overhead logarithmic in the dimension $d$. 
\end{abstract}

\maketitle



\section{Introduction}
In this article we study the one parameter family of $\alpha$-Renyi entropies \cite{renyi1961,MllerLennert2013}: for $\alpha> 0$ and $\alpha\neq 1$, the $\alpha$-Renyi entropy of a quantum state represented by a positive semidefinite operator $\rho\in\C^{d\times d}$, called its density matrix, is defined by
\begin{equation}
    \label{eq:Renyi_entropy}
    S_\alpha(\rho):=\frac{1}{1-\alpha}\log\left[\Tr{\rho^\alpha}\right].
\end{equation}
Taking the limit $\alpha\to 1$ gives the familiar von Neumann entropy, $S(\rho):=-\Tr{\rho\log\rho}$. Classical (discrete) probability distributions can be subsumed into this notation by considering a probability mass function $p=(p_1,\ldots,p_d)$ to be a density matrix that is diagonal in the computational basis, as $\rho_p =  \text{diag}(p_1,\ldots,p_d)$. $S(\rho)$ reduces to the Shannon entropy when $\rho$ is such a `classical state'.

The notion of entropy has played a key role in the development of a variety of scientific disciplines, ranging from thermodynamics to information theory. It gives us a way to quantify the idea of disorder in a system, and the famous second law of thermodynamics essentially states that the entropy of a closed system can never decrease. A variety of entropic functionals have operational meanings in information theory, and are closely related to the rates at which input data can be transmitted over communication channels.

Since the Renyi entropy is a generalised entropic measure and includes the von Neumann entropy as a special case, it is a problem of interest to estimate the Renyi entropy of unknown states or classical probability distributions, for different values of $\alpha$. Such estimates are found to be useful, for instance, in quantifying the efficacy of an ergodic source as a random number generator \cite{Kim2018}, and in the analysis of network structure, clustering, and signal processing of streams of high-frequency data \cite{Clifford2013}. Furthermore, as shown in \cite{Zhao07}, the Shannon entropy can be estimated to any desired precision by interpolation using estimates of $S_{\alpha}(\rho)$ for values of $\alpha\in(0,2]$.

Entropy functions are also important quantities characterising a quantum system.  In entanglement theory, they can give a measure of the amount of entanglement contained in  bipartite quantum systems -- particularly important in this regard is the Renyi entropy for $\alpha=2$, which is known as the entanglement entropy \cite{Calabrese2009}. Entropic quantities are also often used as operational measures in quantum information-processing tasks \cite{Konig2009}.  As one of the most famous examples, they provide the asymptotic lower bound for compressing quantum data in a noiseless fashion, i.e. Schumacher's noiseless compression \cite{PhysRevA.51.2738}. 

In recent decades, entropy functions have also found intriguing applications in condensed matter physics \cite{Laflorencie2016}, and high energy physics. They have even had immense theoretical implications in the theory of gravity and black holes \cite{Bekenstein1973,Dong2016}, and their study from a quantum information theoretic viewpoint continues to be a rich source of new physical insights \cite{Azuma18,Azuma20}.

Thus it stands an important question to compute the value of these entropy functions efficiently on unknown states. In particular, given access to several copies of a quantum state, how many measurements are required to obtain estimates of a chosen entropy function of the state, to within a desired additive or multiplicative precision? Additionally, if one has access to the dynamic process that prepares the state, in the form of a unitary circuit on a larger system (the purification), does this lead to any improvement in our ability to estimate its entropies?

\subsection{Related work}
We can group studies of entropy estimation into four categories: (1) classical and (2) quantum algorithms for estimating entropies of classical distributions; and (3) classical and (4) quantum algorithms for estimating the entropies of quantum states. There are several studies of the first kind in classical information theory \cite{Batu2002,WU2014Shannon,JiantaoJiao2015,ValiantValiant2011}. 

Coming to the third category, \citeauthor{Hastings2010MeasuringSimulations} \cite{Hastings2010MeasuringSimulations}, for instance, discuss a quantum Monte Carlo method to measure the $2$-Renyi entropy of a many-body system by evaluating the expectation value of a unitary swap operator. Their method uses a number of samples that scales polynomially number in the system size. 

More in the flavour of quantum algorithms, and in a sense straddling categories (2) and (4), \citeauthor{Acharya2017} \cite{Acharya2017} study the sample complexity of estimating von Neumann and Renyi entropies of mixed states of quantum systems, in a model where as input one gets $n$ independent copies of an unknown $d$-dimensional density matrix $\rho$. They allow arbitrary quantum measurements and classical post-processing, and show that in general the number of quantum samples required scales as $\Theta(d^2/\epsilon^2)$, which is asymptotically the same as the number of samples that would be required to learn the state completely via tomography methods. The experimental measurement of the entropy of specific quantum systems has also recently been investigated \cite{Islam2015}. 

While it enables a tight characterisation of the sample complexity of the problem (table \ref{tab:acharya17}), other potentially stronger input models are also possible which are not captured in this picture. In this paper, we consider an oracular input model that is popular in quantum query algorithms, wherein data is accessed in the form of a quantum state. This state may be the output of some other quantum subroutine, in which case that subroutine itself is the oracle. Such input models can capture the fact that we have access to the process generating the unknown state, which we may \emph{a priori} expect to be useful in reducing the effort required in estimating its properties. 

In this vein, and bringing us to quantum algorithms for estimating the entropies of quantum states (which as noted before subsumes the case of classical probability distributions),  \citeauthor{Li2019} \cite{Li2019} study how to obtain additive approximations to von Neumann and Renyi entropies in an oracular model 
and present upper and lower bounds on the query complexity. \citeauthor{Gilyen2019} \cite{Gilyen2019} study another similar oracular model, known as the `quantum purified query access' model which essentially provides a pure state, sampling from which reproduces the statistics of the original mixed state, or target classical distribution. They obtain the best upper bounds known in the literature, showing that the von Neumann entropy can be estimated with query complexity $\tilde{\O}(\sqrt{d}/\epsilon^{1.5})$ and $\tilde{\O}(d/\epsilon^{1.5})$ respectively for classical distribution and quantum density matrices. Both these papers use quantum amplitude estimation (QAE) \cite{Brassard2002} as the means to estimate the target quantities. However, QAE requires full-fledged fault tolerant quantum computers and may not be available in the near future.

Approximation algorithms that estimate a quantity to within a multiplicative factor (i.e.\ such that the estimate $\tilde{x}$ lies in the interval $[\nicefrac{x}{\gamma}, \gamma x]$ for some $\gamma>1$) are particularly valuable when the target quantity might be small, and these algorithms are often harder and more complex. \citeauthor{Batu2002} \cite{Batu2002} consider the estimation of Shannon entropy to multiplicative precision, showing that $\O(d^{\nicefrac{(1+\eta)}{\gamma^2}}\log d)$ samples suffice to estimate it to within a factor $\gamma$, for classical distributions with $S(p)>\nicefrac{\gamma}{\eta}$. This is almost matched by a lower bound of $\Omega(d^{\nicefrac{(1-\eta)}{\gamma^2}}\log d)$ later proven in \cite{Valiant2011}. To the best of the authors' knowledge, the problem of estimating entropies to within a multiplicative factor has not been considered in the quantum algorithms literature. 


\section{Main Results}

In this article, we consider the estimation of Renyi entropies in the purified quantum query access model, and approach the problem using sampling via a $\cc{DQC1}$ method, rather than using the QAE algorithm.  While being less powerful than quantum amplitude estimation, such sampling techniques have the advantage of requiring less stringent quantum resources. In particular, QAE requires long coherence times, and the application of powers of the input oracle and its inverse conditioned on large ancillary registers. Sampling methods in general trade away these requirements for a quadratic increase in the scaling with the precision parameter $\epsilon$. 

Using a recent iterative method of \citeauthor{Chowdhury2019partitionFuncDQC} \cite{Chowdhury2019partitionFuncDQC}, we show how the trace of power functions of the input state can be estimated to within a suitable multiplicative precision, in order to obtain {additive} approximations of the Renyi entropy. This iterative algorithm has an expected asymptotic runtime that depends on the unknown quantity being estimated. Thus we can obtain better bounds on its complexity than by considering only the worst case asymptotic runtime. 

Our approach is to construct a unitary that encodes (or probabilistically implements) the matrix function $\rho^\alpha$. Then we can estimate its normalised trace using the $\cc{DQC1}$ model. We will assume that $\alpha$ is a constant, and leave it out of complexity considerations. Our first result is an algorithm that outputs an additive approximation to the $\alpha$-Renyi entropy of an unknown quantum state, for $\alpha\neq 1$.
\begin{theorem}
\label{thm:main-renyi}
    Given a unitary process $U_\rho$ on $\C^{d+a}$ which produces a purification $\ket{\uppsi_\rho}$ of a mixed state $\rho\in \C^{d\times d}$ with $\id/\delta\preceq\rho\preceq\id$, for $\alpha>0$ with $\alpha\neq 1$, there exists an iterative quantum algorithm that outputs an estimate $\tilde{S}$ such that
    \[
    \Abs{\tilde{S}-S_\alpha(\rho)} \leq \epsilon,
    \] with high probability. The algorithm runs for at most $\O(\log d)$ rounds, making an expected number $\O(\nicefrac{1}{(x\epsilon)^2})$ of independent applications of a quantum circuit which coherently invokes $U_\rho$ and $U_{\rho}^{\dagger}$ a total of $m=\O(\nicefrac{1}{\delta}\log\nicefrac{d}{\epsilon})$ times, in each case measuring a single output qubit. Here, $x=\frac{1}{d}\!\Tr{\rho^\alpha}$. The algorithm uses $\O(m)$ additional $1$- and $2$-qubit gates, and $\lceil\log d\rceil+2$ ancillary qubits.
    
    Furthermore, when $\alpha\neq 1$ is an integer, $m=\alpha$, making the circuit depth is independent of the dimension $d$.
\end{theorem}
We construct the algorithm proving this claim by using the technique of block-encodings and quantum singular value transformations \cite{Chakraborty2018TheSimulation,Gilyen2019SVD} to implement unitaries that are block encodings of the power functions $\rho^\alpha$ on the system adjoined with ancillary registers, and subsequently estimating the trace of these unitaries in the $\cc{DQC1}$ or ``one-clean qubit'' model of computation \cite{Knill1998} in combination with the method of \cite{Chowdhury2019partitionFuncDQC}. In contrast, in \cite{Gilyen2019} a block encoding of $\log\rho$ is applied to a suitable input state, resulting in a state that encodes the von Neumann entropy as the amplitude of a computational basis state, which is then estimated using QAE. \\

We also consider the problem of estimating the entropy to multiplicative precision by using the same iterative subroutine again, improving on the additive estimate obtained, to arrive at an estimate $\tilde{S}_\alpha$ satisfying $(1-\epsilon_{rel})S_\alpha\leq \tilde{S}_\alpha\leq (1+\epsilon_{rel})S_\alpha$, with an overhead that is at most logarithmic in the dimension of $\rho$.
\begin{theorem}
\label{thm:main-renyi-mult}
    Iterating the algorithm of Theorem \ref{thm:main-renyi}, we can obtain an estimate $\hat{S}$ which satisfies
    \[
    \Abs{\frac{\hat{S}}{S_\alpha(\rho)}-1}\leq\epsilon_{rel}
    \]
    with high probability. The algorithm runs for at most
    \[
        R = \O\left(\log\left(\frac{\log d}{\delta}\right)\right)
    \]
    rounds, and the total expected number of runs of the $\cc{DQC1}$ circuit and corresponding single qubit measurements is given by
    \[
        \O\left(\frac{1}{\left(\epsilon x S_\alpha(\rho)\right)^2}\right),
    \]
    where as before, $x=\frac{1}{d}\!\Tr{\rho^\alpha}$.
\end{theorem}
Since by our assumption and the definition of the entropy we have $\O(\delta)\leq S_\alpha(\rho)=\frac{1}{1-\alpha}\log dx$, we note that the the factor $\nicefrac{1}{S_\alpha(\rho)}$ is large only when $dx\to 1$, so that this is at most a factor of $\nicefrac{1}{\delta^2}$ worse than the expected number of measurements required in Theorem \ref{thm:main-renyi}. This problem of approximating the entropy of a state to within a multiplicative precision, which is generally more difficult than additive approximation, has not been discussed in the quantum algorithm literature to the best of our knowledge. We provide a comparison between our work and some of the known results in Table~\ref{tab:acharya17}. \\

We thus extend the investigation of evaluating entropy functions to the case of Renyi entropy, in the purified quantum query access model considered in \cite{Gilyen2019}. Our key contributions are: (1) the use of unitary block encodings of the target operator functions obtained using quantum matrix function implementation techniques, in combination with (2) the replacement of QAE with trace estimation using one clean qubit, and (3) obtaining approximations to multiplicative precision. Since it uses the one-clean qubit model, our method does not require long coherence times or high circuit depth. Furthermore, only a single clean and well-controlled qubit is required, while the remainder can start off in the maximally mixed (highly noisy) state; apart from being a low resource requirement, this also makes our algorithm potentially feasible for testing on near-term NMR or linear optics based quantum hardware. Finally, our runtime analysis using the algorithm recapped in Appendix \ref{app:multiplicative} allows us to bound the expected number of measurements as a function of the unknown target quantity, offering better bounds than would be given by just a worst case analysis.

\begin{table*}[htb]
    \renewcommand{\arraystretch}{1.5}
    \centering
    \begin{tabular}{|c|c|c c|}
        \toprule
         $S_\alpha$  & Copies of $\rho$ ($\Theta(\cdot)$) & $\Exp{\text{\#mmts}}$ & Queries to $U_\rho$ per use of circuit\\ \midrule
         
         $\alpha<1$ & $\left(\nicefrac{d}{\epsilon}\right)^{2/\alpha}$ & $\O(\nicefrac{1}{(x\epsilon)^2})$ & $\O\left(\frac{1}{\delta}\log\frac{d}{\epsilon}\right)$\\
         $\alpha>1$ non-integer & $\left(\nicefrac{d}{\epsilon}\right)^{2}$ & $''$ & $''$\\      

         $\alpha>1$ integer & $\frac{d^{2-2/\alpha}}{\epsilon^{2/\alpha}}$ & $''$ & $\O(\alpha)$\\
         \bottomrule
    \end{tabular}
    \caption{Sample complexity from \cite{Acharya2017} for estimating the Renyi entropies of an unknown $d$-dimensional mixed state to additive precision $\epsilon$ for different ranges of $\alpha$, contrasted with expected number of measurements in this paper (Section \ref{sec:algorithm}). In the worst case, these scale as $\O(d^2)$ for $\alpha<1$, and $\O(d^{2\alpha})$ for $\alpha>1$. (We use $''$ to indicate the same expression as on the preceding line)
    }
    \label{tab:acharya17}
\end{table*}


\section{Preliminaries}
\subsection{Input model}
We assume access to a unitary process $U_\rho$ on $\C^{d}\otimes\C^{a}$ which produces a purification $\ket{\uppsi_\rho}$ of the actual input state $\rho$ in $\C^{d\times d}$
\begin{equation}\label{eq:purified-access-input-oracle}
    U_\rho\ket{0} = \ket{\uppsi_\rho} = \sum_{i=1}^n\sqrt{p_i}\ket{\psi_i}_d\ket{\phi_i}_a,
\end{equation}
so that $\Tr[a]{\ketbra{\uppsi_\rho}{\uppsi_\rho}}=\rho$. The $\{\ket{\phi}_a\}$ and $\{\ket{\psi}_d\}$ are sets of orthonormal vectors on the ancillary and system subspaces respectively. This model, known as the purified quantum query access model, is also discussed by \cite{Gilyen2019} and \cite{Belovs2019} in the context of property testing. 

Note that the case of a classical probability distribution on $d$ points with sampling access is subsumed into this model by embedding it into the diagonal state $\rho_p = \sum_{i=1}^d p_i\ket i$.

\subsection{Implementing power functions of Hermitian matrices}
A block-encoding $U_A$ of a Hermitian matrix $A$ is essentially a unitary that encodes a (sub-)normalised version of $A$ in its top left block, i.e.\
\begin{equation}\label{eq:block-encoding}
    U_A=\begin{pmatrix}
    A/\Lambda & \cdot \\
    \cdot & \cdot
\end{pmatrix},
\end{equation}
where $\Lambda\geq \norm{A}$. The behaviour and use of such encodings has been explored extensively in the last few years \cite{Low2016,Chakraborty2018TheSimulation,Gilyen2019SVD}. Given access to $U_A$, a variety of smooth matrix functions (defined on the spectrum of A) may be implemented, in the sense that a new block encoding $U_A^f$ can be obtained such that
\begin{equation}\label{eq:block-encoding-function}
    U^f_A=\begin{pmatrix}
    f(A)/\beta & \cdot \\
    \cdot & \cdot
\end{pmatrix},
\end{equation}
where $\beta\geq\norm{f(A)}$. In particular, here we are interested in power functions $f(x)=x^\alpha$ for an exponent $\alpha>0$. These can be realised using e.g.\ Lemma 9 of \cite{Chakraborty2018TheSimulation} or Corollary 67 of \cite{Gilyen2019SVD}, with minor modifications. Assuming that $\rho$ not a pure state, and has minimum eigenvalue $\delta\in(0,\nicefrac12)$, an $\epsilon$-approximate block encoding of $\rho^\alpha$ can be created with $\O(\frac {1}{\delta}\log{\frac 1\varepsilon})$ uses of $U_\rho$. Even if $\rho$ has a non-trivial kernel, it is fairly easy to implement the matrix function only on the non-singular part of the input (e.g.\ \cite{Harrow2009QuantumEquations,Gilyen2019}), and for classical distributions, we can consider the restriction to the support of the distribution by pre-processing using e.g.\ sparse PCA.

The precision $\varepsilon$ specifies how close the top left block of the new encoding is to $\rho^\alpha$ in the operator norm. For integral values of $\alpha$ we can obtain an exact encoding with $\varepsilon=0$, e.g.\ using Chebyshev polynomial methods as in \cite{subramanian2018}; this has the effect of removing the logarithmic factors from the complexity for integral $\alpha$. We collect the necessary results about implementing matrix function in Appendix \ref{app:matFunc}.



\subsection{$\cc{DQC1}$ Model }
\label{sec:dqc1}

The one clean qubit or $\cc{DQC1}$ (for \emph{Deterministic Quantum Computation with one clean qubit}) paradigm was inspired by studies of NMR based quantum computing \cite{Schulman99NMR}. The model generated interest due to the potential for physical implementation, since only a single `clean' qubit has to be initialised to a pure state, while the remaining `dirty' qubits can start in highly mixed (or random) states. The possibility that even such a weak model can be used to perform useful computations is striking \cite{Knill1998,Jordan2008Thesis}, because if all the qubits are in the maximally mixed state no non-trivial computation can be done (no unitary process can purify the state). Algorithms in the $\cc{DQC1}$ model always have a fixed initial state of $n+1$ qubits, and both the problem instance and algorithm are encoded into a poly$(n)$-sized unitary circuit to be applied on them. The answer is encoded into the probability of obtaining the outcome zero on measuring the clean qubit at the end of the computation. The problem of computing the normalised trace of a unitary was shown to be complete for $\cc{DQC1}$ \cite{Knill1998}. Further work indicated the presence of non-classical correlations in $\cc{DQC1}$ computations \cite{Datta08DQC}. More recently, it was shown that it is classically hard to sample from the output distribution of a $\cc{DQC1}$ computer up to multiplicative error, conditional on standard complexity theoretic assumptions \cite{Morimae14DQC1hardness,Fujii18DQC1}.

The $\cc{DQC1}$ or ``one-clean qubit'' model of computation is based on the use of a single well-controlled or `clean' qubit, and a number $n$ of noisy qubits that are taken to be in the maximally mixed state \cite{Knill1998,Shor2007}. Algorithms in this model are embedded into some controlled $n$-qubit unitaries, and the outputs are encoded into the probability of observing $0$ on measuring the clean qubit. Estimating the normalised trace of a unitary is known to be a $\cc{DQC1}$-complete problem \cite{Knill1998}.
\begin{figure}[htbp]
\begin{center}
\[
    \Qcircuit @C=0.5em @R=0.5em {
    & &\lstick{\ket{0}}&\gate{H} & \ctrl{1} & \gate{H} & \meter \\
    & \push{~~~~~}&\lstick{\frac{\id}{2}}&\qw &  \multigate{3}{~U~} & \qw & \qw \\
&&\lstick{\frac{\id}{2}}&\qw & \ghost{~U~} & \qw & \qw &   \\
&&\push{\vdots} && & \push{~\vdots}& &&   \\
&&\lstick{\frac{\id}{2}}&\qw & \ghost{~U~} & \qw & \qw &
}
\]
\end{center}
    \caption{A $\cc{DQC1}$ circuit that can be used to estimate $\nicefrac{1}{2^n}\!\Tr{U}$, for which no classical efficient algorithm is known. Measurements are made in the computational basis.
    } 
\label{fig:dqc1_trace}
\end{figure}
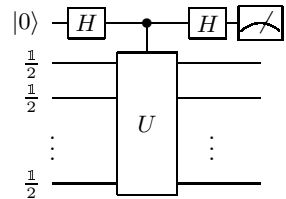

The initial state consists of one qubit set to the $\ket 0$ state, and $n$ qubits in the maximally mixed state, i.e.\ $\xi_{\text{in}} = \ketbra{0}{0}\otimes\id_n/2^n=\frac{\id+Z}{2}\otimes\id_n/2^n$. We can write the final state in figure \ref{fig:dqc1_trace} after the application of the circuit but before measurement as
\begin{align}
    \xi_{\text{out}}
        &=\frac{1}{2^{n+1}}\begin{pmatrix}
            \id_n & U^{\dagger} \\
            U & \id_n
        \end{pmatrix},
\end{align}
from which we see that the expectation values of the Pauli $X$ and $Y$ operators for the clean qubit give the estimates $\langle X\rangle = 2^{-n}\text{Re}(\Tr{U})$ and $\langle Y\rangle = -2^{-n}\text{Im}(\Tr{U})$. These can be extracted to within additive precision $\epsilon$ using $\O(\log\frac{1}{\eta}/\epsilon^2)$ measurements, with probability at least $1-\eta$ for $\eta>0$. Since the dependence on $\eta$ is logarithmic and the success probability can easily be boosted by repetition or other standard methods, we will drop the factor in $\eta$ from further consideration.

As discussed in \cite{Jordan2008Thesis}, and used for example by \citeauthor{Cade18} \cite{Cade18}, the $\cc{DQC}k$ model with $k$ clean qubits can be used to obtain the trace of a submatrix whose size is an inverse-poly sized fraction of the whole unitary. This is useful in our context, because we deal with a unitary block encoding of matrix functions such as $\rho^\alpha$, so that these target matrices (whose trace we are interested in) are submatrices located in the top-left corner of the unitary.

\section{Proof of Theorem \ref{thm:main-renyi}}
\label{sec:algorithm}
In order to prove Theorem \ref{thm:main-renyi} by describing the algorithm and analysing its complexity, we begin by considering how the error in estimating $x=\Tr{\rho^\alpha}$ propagates to the estimate of $S_\alpha(\rho)$ calculated using $x$.

\subsection{Error analysis}
\label{sec:error-analysis}
From the definition of the $\alpha$-Renyi entropy (Eqn.~\eqref{eq:Renyi_entropy}), if we use an estimate $\tilde{x}$ of $x=\Tr{\rho^\alpha}$, the corresponding error in $S_\alpha(\rho)$ is given by
\begin{align*}
    \Abs{\tilde{S}_\alpha(\rho)-S_\alpha(\rho)}&=\Abs{\frac{\log\tilde{x}}{1-\alpha} - \frac{\log x}{1-\alpha}} \nonumber\\
    &= \Abs{\frac{1}{1-\alpha}\log\frac{\tilde{x}}{x}}.
\end{align*}
From the above expression it is clear that if $\tilde{x}$ an estimate of $x$ to a multiplicative factor $\gamma>0$, satisfying
\begin{equation*}
    \frac{x}{\gamma}\leq \tilde{x}\leq \gamma x,
\end{equation*} 
then we obtain an estimate $\tilde{S}_\alpha(\rho)$ to an additive precision given by
\[
    \Abs{\tilde{S}_\alpha(\rho)-S_\alpha(\rho)} \leq \Abs{\frac{\log\gamma}{1-\alpha}}
\]
For `tight' multiplicative approximations, with $\gamma=1+\epsilon_{rel}$ where we shall refer to $\epsilon_{rel}<1$ as the 
multiplicative precision, we can write
\begin{equation}
    (1-\epsilon_{rel})x\leq \tilde{x}\leq (1+\epsilon_{rel})x.
\end{equation}
For such $\gamma$, since $y-\frac{y^2}{2}\leq\log (1+y)\leq y$, we have
\begin{equation}
    \Abs{\tilde{S}_\alpha(\rho)-S_\alpha(\rho)} \leq \frac{\epsilon_{rel}}{\Abs{1-\alpha}}.
\end{equation}

Thus, an upper bound on the complexity of estimating $S_\alpha(\rho)$ to additive precision $\epsilon$ is directly given by that of the method used to estimate $\Tr{\rho^\alpha}$ to multiplicative precision $\epsilon|1-\alpha|$. 

Let us now consider how to obtain approximations of the (normalised) trace of $\rho^\alpha$ using the $\cc{DQC1}$ method described in the previous section.

\subsection{Estimating $\Tr{\rho^\alpha}$}
From Appendix \ref{app:matFunc} we first use \cref{lem:rho_block_encoding} to obtain the block encoding of $\rho$, and then use \cref{lem:powerfunctions} to construct an $\varepsilon$-approximate block encoding for $\rho^\alpha$:
\[
U_\rho = \begin{pmatrix}
    \rho & \cdot \\
    \cdot & \cdot
\end{pmatrix} \mapsto \begin{pmatrix}
    f_\alpha(\rho) & \cdot \\
    \cdot & \cdot
\end{pmatrix} =: U_{\alpha},
\]
where $\|f_\alpha(\rho)-\rho^\alpha\|<\varepsilon$. The unitary $U_{\alpha}$ requires $\O\left(\frac{\max(1,\alpha)}{\delta}\log\frac 1\varepsilon\right)$ uses of the block encoding of $U_\rho$, and two more than the number of ancillary qubits as $U_\rho$, where $\delta>0$ lower bounds the least eigenvalue of $\rho$. Since the dependence of the complexity on $\varepsilon$ is logarithmic, we can afford to choose exponentially small $\varepsilon$ if necessary, incurring only a polynomial overhead. In fact, we shall choose $\varepsilon=\epsilon/d$, since the error in the trace is then bounded by $d\varepsilon=\O(\epsilon)$. This will only result in a factor of $\O(\log\nicefrac{d}{\epsilon})$ in the query complexity.

The normalised trace of this $d+a$-dimensional unitary $U_\alpha$ can be estimated to additive precision $\epsilon$ with high probability by making $\O(\log\nicefrac{1}{\epsilon^2})$ uses of the unitary (or more precisely of the $\cc{DQC1}$ circuit in Fig. \ref{fig:dqc1_trace}). Since $U_\alpha$ has the block form
\begin{equation*}
    U_\alpha=\begin{pmatrix}
    \rho^\alpha & \cdot \\
    \cdot & \cdot
\end{pmatrix},
\end{equation*}
its trace contains a contribution from $\Tr{\rho^\alpha}$. What we would like is to isolate this term alone. Importantly, we know that $\Tr{\rho^\alpha}$ will be real and positive.\\

As mentioned in Section \ref{sec:dqc1} this can be done, for example, by choosing to measure the ancillary qubits of $U_\alpha$, and requiring them to be in the $\ket0$ state, hence projecting onto the relevant subspace to capture the trace of the submatrix $\rho^\alpha$.  We also describe another simple way of doing the same task below. Consider multiplying $U_\alpha$ by a controlled phase operator to convert the block-encoded matrix alone to $i\rho^\alpha$, i.e. consider the unitary $U'_\alpha=U_\alpha V$ where the unitary $V$ applies a conditional phase of $e^{i\pi/2}=i$,
\begin{equation}
    V=i\ketbra{0}{0}\otimes\id + \sum_{k=1}^{a-1}\ketbra{k}{k}\otimes\id,
\end{equation}
which can easily be arranged using an ancillary qubit initialised to the $\ket{+}$ state to which a conditional rotation of $R_y(\pi/2)$ is applied. Now, exploiting the fact that $\Tr{\rho^\alpha}$ is purely real, we can estimate it by $\text{Re}(\Tr{U_\alpha})-\text{Re}(\Tr{U'_\alpha})$, using only twice as many measurements and uses of $U_\alpha$ as required for estimating $\text{Re}(\Tr{U_\alpha})$ itself. 

Finally, recall that with $\cc{DQC1}$ we get the \emph{normalised} trace. However as discussed in Section \ref{sec:error-analysis}, since we are interested in a multiplicative approximation to the trace, this does not pose any issue --- approximations to within a multiplicative factor are unaffected when scaled by a constant. Thus an estimate of $\frac{1}{d}\!\Tr{\rho^\alpha}$ to multiplicative precision $\epsilon$ is also a valid estimate of $\Tr{\rho^\alpha}$ to within the same multiplicative precision.

\subsection{Estimating the trace to multiplicative precision}

In Appendix \ref{app:multiplicative}, we review an iterative method introduced in \cite{Chowdhury2019partitionFuncDQC} for improving an additive precision estimate to a multiplicative one, by starting off with inaccurate additive estimates (with large $\epsilon$) and progressively improving the precision, without requiring too many iterations. Combining this idea encapsulated in Algorithm \ref{alg:multiplicative} with the estimation of normalised trace, we obtain an estimate $\tilde{x}$ that satisfies
\[
\Abs{\frac{\tilde{x}}{x}-1}<\epsilon_{rel}.
\]
It is worth mentioning that the only valid multiplicative approximation to \textit{any} precision of a quantity that takes value zero is zero itself. We do not need to worry about the possibility of $x=0$ here, because the trace of $\rho^\alpha$ is non-zero for all quantum states $\rho$ and finite values of $\alpha$. 

From the discussion in Appendix \ref{app:multiplicative}, we conclude that at most 
\[
\O\left(\frac{x_{\max}}{x_{\min}}\right)
\]
iterations are required, and the total expected number of uses of the block encoding of $\rho^\alpha$ and subsequent single qubit measurements is given by 
\[
\O\left(\frac{1}{\left(x\epsilon\right)^2}\right),
\]
where $x=\frac1d\!\Tr{\rho^\alpha}$ is the unknown quantity being estimated. The minimum and maximum values of the normalised trace are straightforward to compute, and we discuss them below for completeness.

\subsection*{Minimum and maximum values of $\Tr{\rho^\alpha}$}
\label{sec:min-max-trace}

Since we know states are normalised so that $\Tr{\rho}=1$, the eigenvalues $p_1,\ldots,p_d$ of $\rho$ form a probability distribution. The minimum and maximum values of the trace of $\rho^\alpha$ can then easily be calculated (e.g. see Lemma 5 of \cite{Acharya2017}). Using the definition of the trace
\begin{equation*}
    \Tr{\rho^\alpha} := \sum_{i=1}^d p_i^{\alpha},
\end{equation*}
we have for $\alpha<1$
\begin{align*}
    1\leq \Tr{\rho^\alpha} \leq d^{1-\alpha},
\end{align*}
and for $\alpha>1$
\begin{align*}
    d^{1-\alpha}\leq \Tr{\rho^\alpha} \leq 1,
\end{align*}

Correspondingly, for the normalised trace we have the ranges
\begin{align}
\label{eq:min-max-trace}
    \frac{1}{d}\Tr{\rho^\alpha}\in\begin{cases}
                    [d^{-1},d^{-\alpha}] & 0<\alpha<1\\
                    [d^{-\alpha},d^{-1}] & 1<\alpha
                \end{cases}
\end{align}

Hence, for the $\alpha<1$ case we can also consider scaling up the block encoding of $\rho^\alpha$ by $d$ using QSVT techniques for ``pre-amplification'' which we recall in Lemma \ref{lem:amplify} of Appendix \ref{app:matFunc}. This will make the normalised trace larger than unity and hence any additive approximation is also a multiplicative approximation to the same precision, whence the cost for estimating the entropy becomes $\nicefrac{1}{\epsilon^2}$ measurements in the $\cc{DQC1}$ of a unitary block encoding circuit of $d\rho^\alpha$ that uses $U_\rho$ and $U_{\rho}^{\dagger}$ around $\O\left(\frac{d\log(\nicefrac{d}{\epsilon})}{\epsilon^2\delta}\right)$.


\section{Proof of Theorem \ref{thm:main-renyi-mult}}
\label{sec:mult-approx}

\subsection*{Minimum and maximum values of $S_\alpha(\rho)$ given $\rho$ has non-trivial eigenvalues}
Suppose the least eigenvalue of $\rho$ is known to be some fixed  $p_{\min}=\delta\in(0,1/2)$. We only consider the case where $\delta<\nicefrac1d$, since otherwise the actual minimum eigenvalue will be zero since the state is normalised. 

With the minimum eigenvalue fixed, the minimum and maximum of the Renyi entropy is achieved respectively by the minimisers and maximisers of the entropy functional for the remaining $d-1$ dimensional distribution; we can simply use the minimum and maximum values of $\Tr{\rho^\alpha}$ from \eqref{eq:min-max-trace} in order to bound these. 

Note that when the density matrix has a non-trivial kernel, the maximum entropy can be $\O(1)$ independent of the dimension. For example if we take $\delta=\nicefrac14$, the uniform distribution over the remaining $d-1$ dimensions is excluded from our consideration for $d\geq 5$. However, we will not need to make such a detailed analysis.

Since the factor of $1-\alpha$ in the definition of the entropy changes sign at $\alpha=1$, we do not need to consider $\alpha>1$ and $\alpha<1$ separately; the minimum and maximum entropy are the same in both cases. Taking the $\alpha>1$ case for illustration, we have in this case $1-\alpha<0$ and $(d-1)^{1-\alpha}<1$, so
        \begin{align}
        \label{eq:trace-bounds-p-min}
            S_{\alpha}^{\min} &= \frac{1}{1-\alpha} \log\left(\delta^\alpha + (1-\delta)^\alpha \right) \nonumber\\
            &\geq \frac{\alpha}{1-\alpha}\log (1-\delta) \geq \frac{\delta\alpha}{|\alpha-1|}\nonumber\\
            S_{\alpha}^{\max} &= \frac{1}{1-\alpha}\log\left(\delta^\alpha + (d-1)^{1-\alpha}(1-\delta)^\alpha\right)\nonumber\\
            &\leq \log d
    \end{align}
    Thus we then have, since $0<\delta<\nicefrac12$,
    \begin{align}
        \frac{S_{\alpha}^{\max}}{S_{\alpha}^{\min}}
        &= \O\left(\frac{\log d}{\delta}\right). 
    \end{align}

Thus, from the analysis of Appendix \ref{app:multiplicative} we see that Algorithm \ref{alg:multiplicative} runs for at most
    \begin{equation}
        R = \O\left(\log\left(\frac{\log d}{\delta}\right)\right)
    \end{equation}
rounds, and the total expected number of runs of the $\cc{DQC1}$ circuit and corresponding single qubit measurements is given by \eqref{eq:exp-num-mmts} with $x=S_\alpha(\rho)$ and with the algorithm of Theorem \ref{thm:main-renyi} playing the role of the additive estimation subroutine. That is
\begin{equation}
    \text{number of mmts.}=\O\left(\frac{1}{\left(\epsilon x S_\alpha(\rho)\right)^2}\right),
\end{equation}
where $x=\frac1d\!\Tr{\rho^\alpha}$.


\section{Discussion}
Having presented our algorithm, we now turn to a more detailed discussion of the existing literature related to our work, and point out some of the advantages, both technical and pedagogical, of our approach. While it is true that our method boils down to combining two well studied algorithmic primitives (namely DQC1 and QSVT), we would like to emphasise its significance below, and stress that this combination has not been studied before to the best of our knowledge. 

The authors in Ref.\ \cite{Cade18} also use matrix function techniques and their results can be straightforwardly extended to Renyi entropies; however they focus on Hamiltonian operators and hence use a rather different input model, suitable for Hamiltonian simulation - in particular, their results only apply to sparse Hermitian matrices, and they work with Hamiltonians specified as a linear combination of local terms. On the other hand, although density operators are special classes of Hermitian matrices, they may be dense and are rarely known as a linear combination of local or sparse terms. The purified access input model we use is fundamentally different and inspired by information theory rather than graph or complexity theory, and is quite natural for dealing with density matrices which may be part of the output of some other quantum process or algorithm. The use of more advanced QSVT techniques allows us to leverage this model to estimate the Renyi entropies of these potentially dense matrices which need not be local in any sense of the term, hence setting our methods apart from those of Ref.\ \cite{Cade18}.
	
Similarly, while the authors in \cite{Chowdhury2019partitionFuncDQC} also use ideas like ours in order to improve additive estimates to multiplicative ones, they do not analyse the expected runtime of the algorithm, due to the fact that their application is to partition functions, which can be unbounded as a function of the system dimension. We give a full analysis of the expected runtime behaviour of the iterative routine we have used, exploiting the strong bounds on Renyi entropies in terms of the system dimension. 

Another paper that directly relates to ours in terms of providing the ability to estimate Renyi entropies is that of \citeauthor{Ekert02}~\cite{Ekert02}, wherein the authors make use of a fixed quantum circuit (for any input) which essentially performs the $\cc{Hadamard}$ or SWAP test \cite{Aharonov08}. This test is also a motivation for the basic working of DQC1 \cite{Jordan2008Thesis}.  The input they use consists of iid copies of the mixed input state, and crucially requires circuits to be jointly applied to multiple iid copies of the state at a time. On the other hand, the purified query access input model we use is very different in that it provides access to the unitary process that prepares the input state, and the matrix function implementation techniques make crucial use of this fact to extract information about the state using single copies at a time. The purified access oracle is a very natural input model when the input state is actually the output of some other quantum subroutine.

The controlled shifts used by \citeauthor{Ekert02}~\cite{Ekert02} can be thought to correspond to the basis monomial functions $\rho^k$ in our case. One disadvantage of their technique is the need for a $k$-fold tensor product of iid copies of the input state to be prepared and acted upon with a cascade of SWAP gates, of total depth $k$. The authors do not provide asymptotic bounds on the copy complexity for different functionals, but a straightforward estimation shows that it will scale inverse quadratically in the precision required, using standard Chernoff bounds. While this scaling in the precision is similar to ours, the scaling of their methods with dimension $d$ is worse; the requirement being $kd$ qubits of memory, and roughly $kd^{2k-1}/\epsilon^2$ copies of the input state, to estimate a Renyi entropy of order $k$.

The techniques of implementing quantum matrix functions via quantum singular value transformations (QSVT) in combination with DQC1 or QAE can be used to estimate a wide variety of (nonlinear) functionals of quantum states by implementing a polynomial approximation of the target function. As a simple example which also illustrates the centrality of Renyi entropies as an important class of nonlinear functions, we note that the monomial functions $x^k$ can be used to write down polynomials approximating any function with bounded derivatives, for instance using the Taylor series.

The combination of matrix function techniques and DQC1 is also pedagogically simple, giving us a new perspective on the sample complexities attained using more sophisticated techniques in \cite{Acharya2017}. It also leads to improvements in some regimes: for instance, for $\alpha<1$, the sample complexity our algorithm requires is $n^2(\nicefrac{1}{\delta})^\alpha$, in comparison to the strictly super-quadratic $n^{\nicefrac{2}{\alpha}}$ that is obtained in \cite{Acharya2017}.\\

\parahead{Experimental work: }Two important recent papers that deal with the estimation of quantum Renyi entropies with a focus on experiments are Ref.\ \cite{Elben2018} and Ref.\ \cite{Brydges2019}. \textbf{\citeauthor{Elben2018} \cite{Elben2018}} consider estimating the entropy of quantum (many-body) states of generic Hubbard and spin models. The key breakthrough they demonstrate over previous works (such as \cite{Ekert02}) is the ability to use measurement statistics obtained from measuring individual copies of the state in order to estimate Renyi $\alpha=2$ entropies via statistical relations between measurements made in random bases. Earlier work required joint measurements on multiple copies of the state. The authors state that the number of measurements (equal to the sample complexity) required is polynomial in the Hilbert space dimension, without giving the explicit asymptotic dependence. 

Similar to \cite{Elben2018}, \textbf{\citeauthor{Brydges2019} \cite{Brydges2019}} focus on $\alpha=2$ Renyi entropy, and also show that measurements statistics of individual copies of the input state in random bases can be used to estimate it, eliminating the need for joint measurements. The authors give evidence to argue that their method works for arbitrary quantum states of up to tens of qubits, without providing formal theoretical proof. One of the improvements they obtain over Ref.\ \cite{Elben2018} is in avoiding an exponential overhead in the classical postprocessing. The number of measurements (and so the sample complexity) in this work is given to be $2^N$ where $N$ is the number of qubits. The authors do not state any asymptotic complexity results.

In comparison to Refs. \cite{Elben2018} and \cite{Brydges2019}, our work is agnostic to the physical nature of the input state and will work for any quantum system that can be accessed via its purified access oracle. Furthermore, we study the exact form of the asymptotic sample complexity for all values of alpha except $\alpha=0,1$. Our method also requires only unitary transformations and measurements applied to individual copies of the state. Since the complexity we attain is quadratic in the Hilbert space dimension, we believe this is certainly at least equal to or better than the polynomial complexity attained in Ref.\ \cite{Elben2018}. Furthermore, we remark that the methods of both the above references apply to pure state circuit model quantum computing, where the entropy measured is that across a particular bipartition of a pure state, whereas our model of choice is DQC1 which intrinsically uses mixed state quantum computing complemented by a negligible amount of pure state resources. Our work is also more general as it deals with estimating any Renyi entropy (except $\alpha=0,1$) to both additive as well as multiplicative precision, describing the asymptotic sample complexity in each case as a function of the system size, target precision, and $\alpha$. \\

\parahead{The experimental feasibility of our work: }The DQC1 model requires essentially no pure state initialisation, since the initial resource is a maximally mixed state and a single qubit in a pure state (in fact, even this qubit can be partially polarised). NMR technologies have been studied for several decades by physicists, and very well developed control techniques exist, so that for example, single qubit gate with error rates as low as $1.3\times 10^{-4}$ were already achieved before 2010 (for an overview, see e.g.\ \cite{Passante2012}). In contrast, similar error rates have only been achieved in the state of the art today in superconducting qubit technology. Furthermore, NMR systems have long coherence times, of the order of seconds, allowing for deep circuits of up to several hundreds of gates to be implemented before decoherence sets in. Similarly, DQC1 achieved with Linear optics technology is believed to be scalable, and has faster readout times, allowing for the order of millions of measurements to be performed per second \cite{Lanyon08LinearOptics}. 

Thus any hopes of testing our work experimentally in the near term only stem from the fact that mixed state resources are generally easier to prepare and mixed state computing is thought to be easier to realise. We do not, however, believe that our method can be implemented on existing devices such as Google’s Sycamore chip. We also emphasise that our focus is on asymptotic sample complexity, rather than implementation, and we do not study the constant factors that affect the exact complexity. Our claims about ease of implementation in comparison to QAE based methods relate specifically to the issue of deep QAE circuits that need only a few measurements, versus our case of shallow circuits coupled with a large number of measurements. This is the same kind of tradeoff of circuit depth for measurements that is achieved by variational algorithms such as VQE for quantum chemistry in comparison to QAE.

    

\section*{Conclusion}
The method we have proposed here is interesting primarily due to the use of the recently developed algorithmic technique of block encodings along with the one-clean qubit model which is believed to be relatively easier to implement than full fledged error corrected quantum computers. The algorithm is certainly not optimal in either the dimension or the precision, and both improving the query complexity and obtaining lower bounds in this model would be interesting. These block encoding methods allow the implementation of several other matrix functions, which may facilitate the estimation of other entropy-like matrix functionals; there are also several possible applications of estimating entropic functionals as a subroutine in algorithmic procedures for pattern matching, compression tasks, and network analysis.

Trace estimation using the $\cc{DQC1}$ method can be motivated by the so-called $\cc{Hadamard}$ test \cite{Jordan2008Thesis}. The advantage in using the $\cc{DQC1}$ method is that only a constant number of well controlled, `clean' qubits are required, and the task of initial state preparation is significantly eased. On the other hand, using the $\cc{Hadamard}$ test to estimate measurement outcome frequencies requires the preparation of suitable initial states which introduces additional sources of error and complexity. Similarly, the amplitude estimation methods which have previously been used for entropy estimation in quantum property testing algorithms require long coherence times and the application of powers of $U_\rho$ and $U_{\rho}^\dagger$ controlled on ancillary registers, leading to deep circuits, and in general need full fledged fault-tolerant quantum computers. 

Using the iterative method of Algorithm \ref{alg:multiplicative} brings the advantage of obtaining approximations to multiplicative precision, with its stopping condition ensuring that the algorithm runs for at most $\O(\log\nicefrac{d}{\delta})$ iterations for states with entropy at least $\delta$. As discussed in Appendix \ref{app:multiplicative}, the expected number of iterations in fact depends logarithmically on the ratio of the unknown target quantity $S$ to its maximum possible value, and consequently the total number of runs of the $\cc{DQC1}$ subroutine is depends quadratically on $S^{-1}$ --- this is particularly valuable since we can explicitly bound the expected runtime by a function of the unknown target quantity, while this kind of analysis is not often possible. Entropies can indeed take values in $[0,1]\subseteq[0,\log d]$ where $d$ is the dimension of the system. It is important to note that it would be useful to have some other method to test if the input state is pure, since then it would have zero entropy and our algorithm for obtaining multiplicative estimates would fail. 

The definition of the complexity class $\cc{DQC1}$ requires that the number of runs or measurements made scales polynomially in the total number of dirty qubits. Thus we clarify that our method, which requires $\epsilon$ to scale inverse polynomially with $d$, does not place (the decision variant of) entropy estimation in $\cc{DQC1}$. However, it would be an interesting question to see if the close connection between entropy estimation and trace estimation can be exploited to solve the problem in $\cc{DQC1}$, or if it would be possible to prove lower bounds showing that it is strictly harder. Entropy estimation commonly falls under the category of problems of distributional property testing. It is worth investigating what kinds of property testing tasks can be solved within $\cc{DQC1}$, and what lower bounds can be proved in this model. Such studies would also throw further light on the power and limitations of this restricted model of quantum computation.\\

\begin{acknowledgements}
The authors would like to thank Hao-Chung Cheng and Johannes Bausch for helpful discussions, and anonymous referees for helpful feedback. SS was supported by a Cambridge-India Ramanujan scholarship from the Cambridge Trust and the SERB (Govt. of India) when the work for this project was carried out. 
\end{acknowledgements}





%

\appendix
\section{Implementing power functions of density matrices}
\label{app:matFunc}
\citeauthor{Gilyen2019SVD} \cite{Gilyen2019SVD} give a series of lemmas showing how to implement block encodings of different kinds of inputs, of which we will be interested in the case of density operators. 

\begin{definition}[Definition 43, \cite{Gilyen2019SVD}]
    An $(\alpha, a, \epsilon)$ block encoding of an operator $A$ acting on $s$ qubits is a unitary $U$ acting on $a+s$ qubits, such that 
    \begin{equation}
        \norm{A - \alpha\Pi^\dagger U\Pi} \leq \epsilon,
    \end{equation}
    where the first register consists of ancillary qubits,  $\Pi:=\ket{0}^{\otimes a}\otimes\id_s$ is an isometry $\Pi:(\C^2)^{\otimes s}\mapsto\text{span}_{\C}\{\ket{0}^{\otimes a}\}\otimes (\C^2)^{\otimes s}$, and $\alpha,\epsilon\in(0,\infty)$.
\end{definition}

The conversion of a purified access oracle as in Eqn.~\eqref{eq:purified-access-input-oracle} into a block encoding for $\rho$ can be achieved using the following result.  

\begin{lemma}[Lemma 45, \cite{Gilyen2019SVD}]
\label{lem:rho_block_encoding}
    Given a unitary $U_\rho$ acting on $a+s$-qubits, which prepares a purification $U_\rho\ket{0}\ket{0}=\ket{\rho}$ of an $s$-qubit density operator $\rho$, such that $\Tr[a]{\proj{\rho}}=\rho$, the unitary
    \begin{equation*}
        U=\left(U_\rho^\dagger\otimes\id_s\right)\left(\id_a\otimes \cc{SWAP}_s\right)\left(U_\rho\otimes\id_s\right)
    \end{equation*} gives an exact $(1, a+s, 0)$ block encoding of $\rho$.
\end{lemma}

This means that the unitary $U$ has the block form 
\begin{equation*}
    U = \begin{pmatrix}
            \rho & \cdot \\
            \cdot & \cdot
        \end{pmatrix},
\end{equation*}
where we have not specified the $a+s-1$ other $s$-qubit blocks on the diagonal.

Such block encodings can be used to implement smooth functions of the input matrix via polynomial approximations, with the following theorem.

\begin{theorem}[Theorem 56, \cite{Gilyen2019SVD}]
\label{thm:QSVT}
    Given an $(\alpha, a, \epsilon)$ block encoding $U$ of a Hermitian matrix $A$, for any degree $m$ polynomial $P(x)$ that satisfies $\forall~x\in[-1,1],~|P(x)|<\nicefrac12$, there exists a $(1, a+2, 4m\sqrt{\nicefrac{\epsilon}{\alpha}}+\delta)$ block encoding $U_p$ of $P(A/\alpha)$.  We can construct $U_p$ using $m$ applications of $U$ and $U^{\dagger}$, a single application of controlled-$U$, and $\O((a+1)m)$ additional $1$- and $2$-qubit gates. A description of the circuit of $U_p$ can be calculated in $\O(\textnormal{poly}(m, \log\nicefrac{1}{\delta}))$ time on a classical computer.
\end{theorem}

Using Theorem \ref{thm:QSVT}, we can implement $\epsilon$-approximate block encodings of power functions $\rho^\alpha$ on the part of the spectrum of $\rho$ that is contained in $[\delta,1]$ for $\delta>0$ by using polynomial approximations. The lower cutoff $\delta$ is necessary because power functions for non-integer $\alpha$ are not differentiable at $x=0$. Monomials for $\alpha=1,2,\ldots$ can be implemented exactly on the entire domain $[0,1]$.


\begin{lemma}[Corollary 67, \cite{Gilyen2019SVD}]
\label{lem:powerfunctions}
    Given an exact unitary block encoding $U$ of a $d$-qubit density matrix $\rho$, that uses $a$- ancillary qubits, we can implement an $\varepsilon$-approximate block encoding of $\rho^\alpha$ for $\alpha>0$ using  $\O(\frac{\max(1,\alpha)}{\delta}\log \frac 1\varepsilon)$ applications of $U$, and $a+2$- ancillary qubits. Here we assume that the spectrum of $\rho$ is contained in $[\delta,1]$.
\end{lemma}

We also quote the following useful theorem that shows how to amplify the singular values of a block encoding.
\begin{lemma}[Theorem 30, \cite{Gilyen2019SVD}]
\label{lem:amplify}
    Given an exact unitary block encoding $U$ of a Hermitian operator $A$, that uses $a$ ancillary qubits, for $\gamma>1$ we can implement a block encoding $U_{\gamma}$ such that every eigenvalue $\lambda_i<\frac{1-\delta}{\gamma}$ of $A$ is amplified to $\gamma\lambda_i$ to multiplicative precision $\epsilon$
    \[
    \Abs{\frac{\tilde{\lambda}_i}{\gamma\lambda_i} - 1} < \epsilon.
    \] 
    $U_{\gamma}$ requires a single ancillary qubit, and $m$ applications of $U$ and $U^{\dagger}$, and $\O((a+1)m)$ additional $1$- and $2$-qubit gates, where 
    \begin{equation}
        m=\O\left(\frac{\gamma}{\delta}\log \frac{\gamma}{\epsilon}\right)
    \end{equation}
     Here we assume that the spectrum of $A$ is contained in $[\delta,1]$. 
\end{lemma}

\section{Obtaining multiplicative approximations using additive approximations}
\label{app:multiplicative}
Suppose we have an algorithm to obtain an $\epsilon$-additive approximation $\tilde{A}$ to some unknown quantity $A$, i.e.\ $|A-\tilde{A}|\leq \epsilon$.
Often we may be interested in an $\epsilon_{rel}$-multiplicative approximation that satisfies $(1-\epsilon_{rel})A\leq\tilde{A}\leq(1+\epsilon_{rel})A$; this is clearly useful when the target quantity could be small, potentially making it difficult to choose an additive precision in advance.
Given a lower bound $0<\lambda\leq|A|$, an appropriate additive precision can be chosen to get a desired precision multiplicative approximation. The complexity of estimating the multiplicative approximation increases by a factor of $\O(\lambda^{-1})$ over that required for additive approximation. If we know independent of the problem size that $|A|>1$, then any $\epsilon$-additive approximation is also a good $\epsilon'<\epsilon$ multiplicative approximation.

\citeauthor{Chowdhury2019partitionFuncDQC}~\cite{Chowdhury2019partitionFuncDQC} have  described a procedure to produce a multiplicative approximation of a quantity $x$ that has known upper and lower bounds $x_{\max}$ and $x_{\min}$ to precision $\epsilon_{rel}$ using a series of additive precision approximations with exponentially increasing precision in the form
\[
\epsilon_r = \frac{\epsilon_{rel}x_{\max}}{2^{r+1}}.
\]
They show that 
\begin{equation}
\label{eq:R}
R = \big\lceil\log\frac{x_{\max}}{x_{\min}}\big\rceil
\end{equation}
iterations suffice to obtain a multiplicative approximation $\tilde{x}$ satisfying
\begin{equation}
\label{eq:multi}
\Abs{\frac{\tilde{x}}{x}-1}<\epsilon_{rel},
\end{equation}
with high probability. Furthermore, their algorithm  has an expected runtime that depends on $\log \left(\frac{x_{\max}}{x}\right)$, which could be significantly better than the worst case if $x$ is close to its maximum value. For a proof that this algorithm, which we recap in Algorithm \ref{alg:multiplicative}, indeed returns an estimate satisfying Eqn.~\eqref{eq:multi}, we refer to \cite{Chowdhury2019partitionFuncDQC}. We shall discuss its expected runtime below, which will be useful in our analysis.

\begin{figure}
\begin{algorithm}[H]
\numberwithin{algorithm}{section}
\renewcommand{\thealgorithm}{\Alph{section}.\arabic{algorithm}}
\caption{Approximating $x$ to multiplicative precision.}
\label{alg:multiplicative}
\begin{algorithmic}
    \State $c' \gets 1-\nicefrac{(1-c)}{R}$\\ \Comment{success probability for additive estimate}
    \State $r \gets 0$
    \Repeat
        \State $\epsilon_r \gets \frac{\epsilon_{rel}}{2^{r+1}}\cdot x_{\max}$ \Comment{precision for additive estimate}
        \State $x_r \gets x_{\max}/2^r$
        \State $\tilde{x}_r \gets \Call{AdditiveEstimate}{\epsilon_r, c'}$\\ \Comment{obtain $\tilde{x}_r~s.t.~\Pr\left(|x_r-\tilde{x}_r|<\epsilon_r\right)>c'$}
        \State $r \gets r+1$
    \Until{$\tilde{x}_r > x_r$} \\
    \Return $\tilde{x}_r$
\end{algorithmic}
\end{algorithm}
\end{figure}

From the choice \eqref{eq:R} and the stopping condition in Algorithm \ref{alg:multiplicative}, the number of iterations is always $1\leq r\leq R$. Without loss of generality, $\exists~ q\geq 1$ such that 
\begin{equation}
\label{eq:q}
\frac{x_{\max}}{2^{q-1}} > x > \frac{x_{\max}}{2^{q}}.
\end{equation}
To bounded the expected number of rounds, consider the case where the algorithm fails to terminate for $r<q+1$. Then for $r\geq q+1$ the probability that the algorithm does not stop at step $r$ is given by
\begin{equation*}
    \Pr\left(\tilde{x}_r < x_r\right) \leq 1-c',
\end{equation*}
where we get the upper bound by noting that since $r\geq q+1\implies x_r>x$, and the estimate $\tilde{x}_r$ is $\epsilon_r$ close to $x$. This requires the mild underlying assumption that $\epsilon_{rel}<2$. 

Hence the net probability that the algorithm terminates at step $r=q+k$ for $k\geq 1$ is at most $(1-c')^{k-1}c'$. If $c'>1/2$, the expected value of $k$ under this geometric distribution is bounded above by $q+2$, and from Eqn.~\eqref{eq:q} we have
\[
q\leq \log_2\left(\frac{2x_{\max}}{x}\right). 
\]
Since we will use the $\cc{DQC1}$ normalised trace estimation technique as the underlying subroutine to obtain the additive estimate, we can also calculate the total expected number of measurements. This will also give us a bound on the number of uses of the input unitary. The $k^{th}$ iteration requires $\O(\epsilon_k ^{-2})$ measurements, and so if the algorithm terminates at step $r$, the cumulative number of measurements scales as
\begin{align*}
    M_{\leq r} &\approx \sum_{k=0}^r \frac{1}{\epsilon_k^2} 
        = \left(\frac{1}{x_{\max}\epsilon_{rel}}\right)^2 \sum_{k=0}^r 4^{k+1}\nonumber\\
        &= \left(\frac{1}{x_{\max}\epsilon_{rel}}\right)^2 \frac{4^{r+1}-1}{3}\nonumber\\
        &= \O\left(\left(\frac{2^r}{x_{\max}\epsilon_{rel}}\right)^2\right).
\end{align*}
Now we can upper bound the expected value of $M_{\leq r}$ under the geometric distribution for the number of iterations; we have
\begin{align*}
    \Exp{M} &\leq \sum_{k=1}^\infty c'(1-c')^{k-1}M_{\leq q+k}\nonumber\\
        &\leq \left(\frac{1}{x_{\max}\epsilon_{rel}}\right)^2 c' \sum_{k=1}^\infty (1-c')^{k-1}4^{q+k} \nonumber \\
        &=4^{q+1}\left(\frac{1}{x_{\max}\epsilon_{rel}}\right)^2\cdot c'\sum_{k=0}^\infty (4(1-c'))^{k}\nonumber\\
        &=\O\left(\left(\frac{2^{q}}{x_{\max}\epsilon_{rel}}\right)^2\right),
\end{align*}
where we make the mild assumption that $c'>\nicefrac34$. Recalling that $\frac{x_{\max}}{2^{q-1}}>x$, we finally have
\begin{equation}
\label{eq:exp-num-mmts}
    \Exp{M} = \O\left(\left(\frac{1}{x\epsilon_{rel}}\right)^2\right).
\end{equation}
Thus, the total expected number of measurements, or equivalently uses of the input unitary, scales quadratically in the inverse of the target quantity.

This is very useful for our problem: as we saw in Section \ref{sec:error-analysis}, knowing $x=\Tr{\rho^\alpha}$ to multiplicative precision $\epsilon$ allows us to obtain $S_\alpha(\rho)$ to additive precision $\epsilon$.


\end{document}